\documentclass[12pt, preprint, showpacs]{revtex4}
\usepackage{graphicx}				
\usepackage[export]{adjustbox}
\usepackage{amsmath}	
\usepackage{mathrsfs}								
\usepackage{amssymb}
\usepackage{gensymb}
\usepackage{braket}
\usepackage{frcursive}
\usepackage{bbold}
\usepackage{bbm}
\usepackage{natbib}
\usepackage[linktocpage=true]{hyperref}
\usepackage{hypernat}

\begin{document}

\title{General form of the renormalized, perturbed energy density via interacting quantum fields in cosmological spacetimes}

\author{Mahmoud Parvizi}
\email{mahmoud.parvizi@vanderbilt.edu}
\affiliation{Department of Physics and Astronomy, Vanderbilt University, Nashville TN 37235, USA}
\pacs{04.62.+v, 11.10.Gh, 11.10.Wx, 98.80.Cq}

\date{\today}

\begin{abstract}
A covariant description of quantum matter fields in the early universe underpins models for the origin of species, e.g. baryogenesis and dark matter production. In nearly all cases the relevant cosmological observables are computed in a general approximation, via the standard irreducible representations found in the operator formalism of particle physics, where intricacies related to a renormalized stress-energy tensor in a non-stationary spacetime are ignored. Models of the early universe also include a dense environment of quantum fields where far-from-equilibrium interactions manifest expressions for observables with substantive corrections to the leading terms. An alternate treatment of these cosmological observables may be carried out within the framework of algebraic quantum field theory in curved spacetime, where the field theoretic model of quantum matter is compatible with the classical effects of general relativity. Here, we take the first step towards computing such an observable. We employ the algebraic formalism while considering far-from-equilibrium interactions in a dense environment under the influence of a classical, yet non-stationary, spacetime to derive an expression for the perturbed energy density as a component of the renormalized stress-energy tensor associated with common proposals for quantum matter production in the early universe. 
\end{abstract}

\maketitle
\tableofcontents
\pagebreak

\section{Introduction\label{INTRO}}

A covariant description of quantum matter fields in the dynamical spacetime of the early universe is essential to proposals for models of baryogenesis and dark matter production. Calculation of cosmological observables involving the interactions of these fields are usually carried out at tree-level in the standard particle physics approach to quantum field theory, i.e. classical Boltzmann equations augmented with thermally averaged $S$-matrix derived interaction rates quantifying particle production in a covariant generalization of a non-stationary spacetime background (see Refs. \cite{KnT,Wei08} for a pedagogical treatment of the standard quatum formulation of kinetic theory in a cosmological setting). However, in a non-stationary Friedmann--Robertson--Walker (FRW) spacetime the lack of time-translation symmetry, among other concerns, makes the applicability of the particle approach during periods of rapid expansion suspect, e.g. there is no notion of a global or preferred vacuum state serving as a basis of the Fock space formulation (see Refs. \cite{Bir82,Par09} for detailed treatment of the strengths and weaknesses of the operator formulation of quantum fields in curved spacetime). In addition, the current paradigm of modeling the origins of observed inhomogeneity of the universe as well as the observed matter content today presupposes that during some earlier period all quantum fields participated in both near and far from equilibrium interactions with respect to a thermal plasma where the standard quantum formalism, as found in Refs. \cite{LeB96,Das97} for instance, can gives rise to appreciable loop-level corrections to the aforementioned interaction rates \cite{Boy05,Ani08,Ham12,HoS15,Dre16}.

An alternate treatment may be carried out within the algebraic formulation of locally covariant quantum field theory as presented, for example, in Ref. \cite{Bru03} (see Refs. \cite{Wal94,Kha15} for a general introduction to the algebraic approach in the context of curved spacetime). This mathematically rigorous formalism is in general useful for clarifying conceptual issues related to and/or providing a foundation for the calculation of observables with traditionally heuristic justifications. In this work, however, we propose a non-traditional application of the formalism inspired by numerical calculations such as those found in Refs. \cite{Deg13,Hac15} where algebraic quantum field theory is employed in computing and characterizing the energy density of a free scalar field propagating in a non-stationary FRW spacetime. In other words, we seek to employ the established algebraic formalism in a concrete numerical calculation of a cosmological observable and not in the traditional pursuit of a rigorous proof of theorem. Though this numerical calculation may be computationally expensive, as compared to the standard formalism, meeting the requirement that cosmological observables be compatible with the semiclassical Einstein equation; i.e. the stress-energy tensor is the expectation value of a quantum state back-reacting on the metric of general relativity, would seem to justify the cost \cite{Hol01,Hol02,Mor03,Hol05}. 

In the algebraic framework finite time intervals are essential to formulating the physical states of interest in FRW spacetimes given the following considerations: (1) time translation invariance does not allow for a unitary, one parameter group of time shift automorphisms on the algebra of observables, hence a two parameter family of automorphisms is required \cite{Oji86,Buc02}, (2) gravitationally induced excitations of quantum matter fields generally accompany non-stationary spacetimes \cite{Par69,Lue90} where the quantum energy densities are only bound from below when smeared along a timelike curve \cite{Haa84,Few00} such that ground states are defined as states of minimal smeared energy along a finite worldline of an isotropic observer \cite{Olb07,Deg10}, (3) observables related to perturbative quantum interactions are generally defined by an algebra generated by a time averaged perturbation in an arbitrarily small yet finite time slice \cite{Chi08}. Furthermore, the usual notions of thermal equilibrium and non-equilibrium dynamics become somewhat ambiguous in FRW spacetimes. For example, the work in Refs. \cite{Buc02S,Oji03,Gra16} suggests observables computed in a manner consistent with the standard formulation of thermal field theory in Minkowski spacetime may serve only as a reference for the properties of the observed state. 

Hence, we take the first step towards probing for corrections to the standard particle physics approach by deriving an expression, via algebraic quantum field theory in curved spacetime, that is at least in principle amenable to numerical calculation, for the renormalized energy density of a free scalar field subjected during a finite time interval to the influence of a perturbative interaction while propagating in a classical yet non-stationary FRW spacetime. In order to derive this expression for the energy density we must begin with the general evolution of the algebraic state over a finite time interval. As there is no full Poincar\'{e} invariance present in our cosmological model, we make use of a two-parameter family of propagators, including a time averaged perturbative interaction, resulting in a method analogous to the Schwinger--Keldysh closed-time-path \cite{Sch61,Kel64}, however extended to non-stationary spacetimes. The state will then encode both the influence of the perturbative interactions as well as renormalization constraints and ambiguities developed in the literature cited above. To this end we allocate Sec. \ref{GMAF} to the development of a general algebraic model of neutral scalars in which we assume a $\Lambda$CDM cosmology with an initial period of inflation per the results found in Refs. \cite{Pla15p,Pla15i}. In Sec. \ref{REDAS}, we derive Eq. (\ref{mainres}) as the main result of this work; i.e. the general form of the expectation value of the renormalized quantum energy density given the influence, during a finite interval of cosmological time, of perturbative quantum interactions and a non-stationary spacetime background. We include in this section an analysis of a cubic interaction as a concrete example. Finally, we discuss this result and future works in Sec. \ref{DISC}.

\section{General Model in the Algebraic Formalism\label{GMAF}}

We consider a theory of a neutral scalar $\phi(x_{\mu})$ on a globally hyperbolic spacetime $(\mathcal{M}_{\Sigma},\mbox{\textbf{g}})$, i.e. a Loretnzian manifold $\mathcal{M}$ with Cauchy surface $\Sigma$ and metric $\mbox{\textbf{g}}$, via the classical free Lagrangian
\begin{equation}
\mathscr{L}_0 = -\frac{1}{2}\bigg(\mbox{g}^{\mu\nu}\nabla_{\mu}\phi\nabla_{\nu}\phi + m^2\phi^2 + \xi R\phi^2\bigg) 
\end{equation}
given $R$ as the Ricci scalar on $\mathcal{M}_{\Sigma}$, $m$ as the field's mass, and $\xi$ its coupling to gravity. Canonical quantization is realized by constructing the Borchers--Uhlmann algebra, a topological $*$-algebra (with unit) defined as
\begin{equation}
\mathcal{A}(\mathcal{M}_{\Sigma}, \mbox{\textbf{g}})  := \mathcal{A}_0(\mathcal{M}_{\Sigma}, \mbox{\textbf{g}})/\mathcal{I}(\mathcal{M}_{\Sigma}, \mbox{\textbf{g}})
\end{equation}
where $\mathcal{A}_0(\mathcal{M}_{\Sigma}, \mbox{\textbf{g}})=\bigoplus_{n=0}^{\infty}\mathcal\mathcal{D}(\mathcal{M}_{\Sigma}^n)$ given $\mathcal\mathcal{D}(\mathcal{M}_{\Sigma}^0)=\mathbbm{C}$, is the free tensor algebra over $\mathcal{D}(\mathcal{M}_{\Sigma})$ as the space of smooth compactly supported densities $f(x_{\mu})$ on $\mathcal{M}_{\Sigma}$ and $\mathcal{I}(\mathcal{M}_{\Sigma}, \mbox{\textbf{g}})$ the $*$-ideal. The free field $\phi(x_{\mu})$ is henceforth denoted by the formal symbol $A_x$. The smeared fields 
\begin{equation}
A(f) = \int_{\mathcal{M}_{\Sigma}}d\mu_{\mbox{g}}\mbox{ }f(x_{\mu})A_x,
\label{conv}
\end{equation}
where $d\mu_{\mbox{g}}$ is the measure on $\mathcal{M}_{\Sigma}$, generate the algebra $\mathcal{A}(\mathcal{M}_{\Sigma}, \mbox{\textbf{g}})$ such that $f \rightarrow A(f)$ is $\mathbbm{R}$--linear and
\begin{eqnarray}
A(f)^* &=& A(\overline{f})\label{bua1}\\
{[}A(f)\mbox{, }A(g){]} &=& iE(f,g)\label{bua2}\\
A(\widehat{K}f) &=& 0\label{bua3}
\end{eqnarray}
$\forall f,g\in\mathcal{D}(\mathcal{M}_{\Sigma})$ and $A(f),A(g)\in\mathcal{A}(\mathcal{M}_{\Sigma}, \mbox{\textbf{g}})$; while $\mathcal{I}(\mathcal{M}_{\Sigma}, \mbox{\textbf{g}})$ is generated by elements including $\widehat{K}f$ and the causal propagator $E:=E^>-E^<$ defined via the unique advanced$(>)$ and retarded$(<)$ fundamental solutions of the Klein--Gordon operator
\begin{equation}
\widehat{K} = (\square_{\mbox{g}} + m^2 + \xi R). 
\label{kge}
\end{equation}

Here, $\mathcal{M}_{\Sigma}:=\mathbbm{R}\times\Sigma_0$ is a spatially flat FRW spacetime with metric written in the familiar form 
\begin{equation}
ds^2 = dt^2 - a_t^2d\Sigma_0^2,
\end{equation}
such that 
\begin{equation}
\square_{\mbox{g}} = \partial_t^2 + 3H_t\partial_t +\frac{\nabla_{\Sigma}^2}{a_t^2}
\end{equation}
and
\begin{equation}
R = 6\bigg(\frac{\ddot{a}_t}{a_t}+\frac{\dot{a}_t^2}{a_t^2}\bigg)
\end{equation}
given $a\mbox{ : }\mathbbm{R}\rightarrow\mathbbm{R}$ as the scale factor and $H_t:= \dot{a}_t/a_t$ as the Hubble parameter. We pass to conformal time $\eta$ via the relation 
\begin{equation}
dt=a_td\eta
\end{equation}
where the metric becomes 
\begin{equation}
ds^2 = a_t^2[d\eta^2 - d\Sigma_0^2].
\end{equation}
and the Klein--Gordon operator of Eq. (\ref{kge}) is rewritten as
\begin{equation}
\widehat{K}_{\eta} = \frac{1}{a_t^2}\bigg[\partial_{\eta}^2 - \vec{\nabla}^2 + a_t^2m^2 + a_t^2\bigg(\xi-\frac{1}{6}\bigg)R\bigg]
\label{kgec}
\end{equation}
This transformation allows for the expansion of the domain of $a_{t}$ into $(-\infty,\eta_0)$ with an asymptotically de Sitter (dS) spacetime $(\widetilde{\mathcal{M}_{\Sigma}},\widetilde{\mbox{\textbf{g}}})$ where $\widetilde{\mbox{\textbf{g}}} = (\mathbf{\Omega}/a_t)^2\mbox{\textbf{g}}$, given $\mathbf{\Omega}\mbox{ : }\mathcal{M}_{\Sigma}\rightarrow\mathbbm{R}^+$, such that $a_{t}=\mbox{exp}\mbox{ }(H_{\Lambda}t\mbox{ })$ for $\eta\in[-\infty,\eta_0]$ corresponds to an early period of inflation with $H_{\Lambda}$ a constant. $\widetilde{\mathcal{M}_{\Sigma}}$ then contains a cosmological past horizon $\mathfrak{J}^{-}$ as a boundary, i.e. a smooth geodescially complete hypersurface diffeomorphic to $\mathbbm{R}\times\mathbbm{S}^2$ at $\eta\rightarrow-\infty$, with coordinates $(v=t+r,\theta,\phi)$ and metric of Bondi form
\begin{equation}
\widetilde{ds}^2|_{\mathfrak{J}^{-}} = 2d\mathbf{\Omega} dv + d\mathbbm{S}^2.
\end{equation}

\subsection{Homogeneous and Isotropic States}

The algebraic states $\omega\mbox{ : }\mathcal{A}\rightarrow\mathbbm{C}$, where $\omega(A^*A)\ge0$  and $\omega(\mathbbm{1})=1$ $\forall A\in\mathcal{A}$ define the $n-$point functions $\omega(A_1A_2...A_n)$. In the case of quasifree states, i.e. the Gaussian states
\begin{equation}
\omega(A_1A_2...A_n) = 
\begin{cases} 
\sum_{X}\prod_{\{i,j\}\in X}\mbox{ }\omega(A_iA_j) & \quad n\text{ even}\\
0 & \quad n\text{ odd}
\end{cases}
\end{equation}
\begin{center}\emph{$X\equiv$ the set of all possible parings $\{i,j\}$ where $i<j$},\end{center}
we require the two--point function $\omega(A_iA_j)$ be of the physically admissible Hadamard form
\begin{equation}
\omega(A_xA_y) = \lim_{\epsilon\downarrow0}\frac{1}{8\pi^2}\bigg[\mbox{ }\frac{U(x_{\mu},y_{\mu})}{\sigma_{\epsilon}(x_{\mu},y_{\mu})}+V(x_{\mu},y_{\mu})\mbox{ log}\bigg(\frac{\sigma_{\epsilon}(x_{\mu},y_{\mu})}{L^2}\bigg)+F(x_{\mu},y_{\mu})\mbox{ }\bigg]
\label{hadpar}
\end{equation}
where the functions $U$, $V$, and $F$ are smooth real--valued bi-distributions and 
\begin{equation}
\sigma_{\epsilon}(x_{\mu},y_{\mu}) \equiv \sigma(x_{\mu},y_{\mu}) + 2i\epsilon[\tau(x_{\mu})-\tau(y_{\mu})] + \epsilon^2,
\end{equation}
with $\sigma(x_{\mu},y_{\mu})$ the signed squared geodesic distance; while $\tau:\mathcal{M}_{\Sigma}\rightarrow\mathbbm{R}$ is an arbitrary time function, and $L$ the length scale. This allows us to extend the 
factored $*$-algebra to $\mathcal{W}(\mathcal{M}_{\Sigma},\mbox{\textbf{g}})$ such that $\mathcal{A}(\mathcal{M}_{\Sigma},\mbox{\textbf{g}})\subset\mathcal{W}(\mathcal{M}_{\Sigma},\mbox{\textbf{g}})$ where renormalization up to mass and curvature ambiguities is carried out by local and covariant Hadamard point-splitting regularization, i.e. Wick products are defined in the coincidence limit
\begin{equation}
\omega(:A_x^2:):=\lim_{y\rightarrow x}[\omega(A_xA_y)-\mathbbm{H}(x_{\mu},y_{\mu})]
\end{equation}
given the purely geometric Hadamard parametrix $\mathbbm{H}(x_{\mu},y_{\mu})$ as the first two terms in Eq. (\ref{hadpar}) and
\begin{equation}
\omega(:A^n(f):):=\int_{\mathcal{M}_{\Sigma}^n}\prod_{i=1}^nd\mu_{\mbox{g}}(x_i)\mbox{ }f(x_1)\delta(x_1,x_2,...x_n):A_1A_2...A_n:.
\end{equation}
Hence, the time ordered products necessary to define perturbative interactions as well as prove the spin-statistics and CPT theorems allow for reliable cosmological observables \cite{Hol05,Hol06,Hol08a,Hol08b,Hol10}. 

Following the formulation of Ref. \cite{Lue90} with explicit constructions found in Ref. \cite{Sch08} the symmetric part,
\begin{equation}
\omega^s(A(f)A(g)) := \frac{1}{2}\bigg[\omega(A(f)A(g)) + \omega(A(g)A(f))\bigg],
\end{equation}
of a quasifree homogeneous and isotropic states in FRW spacetimes is expressed
\begin{equation}
\omega^s(A(f)A(g)) = \int d^3k\int d\eta_x\int d\eta_y\mbox{ }\mathfrak{X}_{\vec{k}}\mbox{ }\bigg\{\overline{X_{\vec{k}}(\eta_x)}X_{\vec{k}}(\eta_y)+X_{\vec{k}}(\eta_x)\overline{X_{\vec{k}}(\eta_y)}\bigg\}\mbox{ }\overline{ \hat{\overline{f}}_{\vec{k}}(\eta_x)}\hat{g}_{\vec{k}}(\eta_y)
\label{hiqs0}
\end{equation}
with, for example, 
\begin{equation}
\hat{f}_{\vec{k}}(\eta_x) = \int \frac{d^3x}{(2\pi)^{3/2}}\mbox{ }f(\eta_x,\vec{x})\mbox{ exp}(-i\vec{k}\cdot\vec{x})
\end{equation}
as the spatial Fourier transform. The mode functions $X_{\vec{k}}(\eta)$ satisfy 
\begin{equation}
\overline{X_{\vec{k}}(\eta)}X_{\vec{k}}(\eta)'-\overline{X_{\vec{k}}(\eta)}'X_{\vec{k}}(\eta) = i
\label{wrn}
\end{equation}
given $X_{\vec{k}}(\eta)'$ as the derivative with respect to $\eta$. Members of the set of unitarily equivalent mode functions satisfying Eq. (\ref{wrn}) are expressed as a Bogoliubov transformation such that
\begin{equation}
X_{\vec{k}}(\eta) = \mathfrak{p}_{\vec{k}}\mbox{ }T_{\vec{k}}(\eta) + \mathfrak{q}_{\vec{k}}\mbox{ }\overline{T_{\vec{k}}(\eta)}
\label{bt1}
\end{equation}
with $|\mathfrak{p}_{\vec{k}}|^2 - |\mathfrak{q}_{\vec{k}}|^2=1$ and $T_{\vec{k}}(\eta)$ an arbitrary reference mode that satisfies the time portion of $\widehat{K}_{\eta}\mbox{ }T_{\vec{k}}(\eta)=0$. Here, $\mathfrak{X}_{\vec{k}}\ge1/2$ is polynomially bounded in $k$ such that equality obtains the pure state while inequality corresponds to the generic mixed state, i.e. the convex combination 
\begin{equation}
\omega^s(A(f)A(g)) = \sum_n \lambda_n\mbox{ }\omega_n^s\mbox{; }\mbox{ }\lambda_n\ge0\mbox{, }\sum_n \lambda_n=1
\label{ccs0}
\end{equation}
of at least two other mixed states $\omega_i^s$ and $\omega_j^s$ such that $\omega_i^s\neq\omega_j^s$. Crucially, in the sense of distributions, we may restrict the free field state to a Cauchy surface of constant conformal time $\eta$ such that Eq. (\ref{hiqs0}) becomes
\begin{equation}
\omega_{\eta}^s(A(f)A(g)) = 2\int d^3k\mbox{ }\mathfrak{X}_{\vec{k}}\mbox{ }|X_{\vec{k}}(\eta)|^2\mbox{ }\overline{ \hat{\overline{f}}_{\vec{k}}}\mbox{ }\hat{g}_{\vec{k}}.
\label{hiqs1}
\end{equation}

Given the fields in our model are real scalers, there is a Gel'fand--Naimark--Segal (GNS)--representation $\pi_{\omega}\mbox{ : }\mathcal{A}\rightarrow\mathcal{T(D)}$, where $\mathcal{T(D)}$ is the Banach space of linear operators on a dense domain $\mathcal{D}$ of the Hilbert space $\mathcal{H}_{\omega}$, with cyclic vector $\Omega_{\omega}\in\mathcal{D}\subset\mathcal{H}_{\omega}$ such that
\begin{equation}
\omega(A)=\braket{\Omega|\pi_{\omega}(A)|\Omega},
\end{equation}
where the irreducible representations $\pi_{\omega}(A)$ are in one-to-one correspondence with the pure algebraic states and contain the usual annihilation and creation operators over $\mathcal{D}$ as the bosonic Fock space over the one--particle space $\mathcal{H}_{\omega}^{(1)}$. However, for more robust models that include interacting fields of perturbative Yang--Mills theory in a general non-stationary spacetime; an equivalent correspondence with $\pm$-helicity one-particle states of the electromagnetic field is not possible \cite{Kha15,Hol08a}. Hence, we continue in the algebraic framework without regard to a Hilbert space representation.

\subsection{Ground States as States of Low Energy\label{GSSLE}}

We now propose generalized ground states from states of low energy (SLE) as put forward in Ref. \cite{Olb07} with explicit constructions in FRW spacetimes found in Refs. \cite{Deg13,Hac15}. Here, we focus on a massive minimally coupled, i.e. $\xi=0$, free scalar field. We remind the reader that the cosmological observables of interest is the  expectation value of the smeared quantum energy density
\begin{equation}
\mathscr{E}_{A(f)}:=\omega(\mbox{ T}_{00}(:A^2(f):)\mbox{ }) 
\end{equation}
consistent with the local and covariant semiclassical Einstein equation 
\begin{equation}
\mbox{R}_{\mu\nu}(x_{\mu}) - \frac{1}{2}R\mbox{ g}_{\mu\nu}(x_{\mu}) = -8\pi G\mbox{ }\omega\bigg(\mbox{T}_{\mu\nu}(:A_x^2:)\bigg)
\end{equation}
where $\mbox{R}_{\mu\nu}$ is the Ricci tensor, $G$ is Newton's constant, and $\omega(\mbox{ }\mbox{T}_{\mu\nu}(:A_x^2:)\mbox{ })$ is interpreted as the expectation value of the free field stress--energy tensor $T_{\mu\nu}$ corresponding to the quantum matter field $A_x$. Though quantum energy densities restricted to a point are not bound from below \cite{Haa84}, those smeared along the worldline of an isotropic observer in FRW spacetimes do have a lower bound when Hadamard states are consider \cite{Few00}.

SLE are then the quasifree pure homogeneous and isotropic states specified by mode functions that minimize the energy density per mode 
\begin{eqnarray}
\mathscr{E}_{\vec{k}}(\eta) &=& \frac{1}{2a_t^4(2\pi)^3}\bigg(|X_{\vec{k}}'(\eta)|^2 - a_tH_t(|X_{\vec{k}}(\eta)|^2)'\nonumber\\
& & +\mbox{ }(k^2+a_t^2m^2+a_t^2H_t^2)|X_{\vec{k}}(\eta)|^2\bigg)
\end{eqnarray}
via the Bogoliubov coefficients of Eq. (\ref{bt1}) such that
\begin{eqnarray}
\mathfrak{p}_{\vec{k}} &=& \mbox{ exp}\bigg(i[\pi-\mbox{arg }c_2({\vec{k}})]\bigg)\sqrt{\frac{c_1({\vec{k}})}{2\sqrt{c_1^2({\vec{k}})-|c_2({\vec{k}})|^2}}+\frac{1}{2}}\\
\mathfrak{q}_{\vec{k}} &=& \sqrt{\frac{c_1({\vec{k}})}{2\sqrt{c_1^2({\vec{k}})-|c_2({\vec{k}})|^2}}-\frac{1}{2}}
\end{eqnarray}
where, for a comoving observer,
\begin{eqnarray}
c_1({\vec{k}}) &:=& \frac{1}{2}\int_{t_i}^{t_f}dt\mbox{ }f^2(t)\bigg\{|X_{\vec{k}}'(\eta)|^2 -a_tH_t(|X_{\vec{k}}(\eta)|^2)'\nonumber\\
& & +\mbox{ }(k^2 + a_t^2m^2 + a_t^2H_t^2)|X_{\vec{k}}(\eta)|^2\bigg\}
\end{eqnarray}
and
\begin{eqnarray}
c_2({\vec{k}}) &:=& \frac{1}{2}\int_{t_i}^{t_f}dt\mbox{ }f^2(t)\bigg\{X_{\vec{k}}'^2(\eta) - a_tH_t[X_{\vec{k}}^2(\eta)]'\nonumber\\
& & +\mbox{ }(k^2 + a_t^2m^2 + a_t^2H_t^2)X_{\vec{k}}(\eta)^2\bigg\}.
\end{eqnarray}
Convolution with the compactly supported function $f(t)$ is then taken over a finite interval of cosmological time, i.e. $t_i,t_f\in I_t\subset\mathbbm{R}$. In what follows we take as our reference 
\begin{equation}
T_{\vec{k}}(\eta):= \frac{1}{\sqrt{2\Omega_{\vec{k}}(\eta)}}\mbox{ exp}\bigg(-i\int_{\eta_0}^{\eta}d\bar{\eta}\mbox{ }\Omega_{\vec{k}}(\bar{\eta})\bigg).
\end{equation}
where
\begin{equation}
\Omega_{\vec{k}}(\bar{\eta})=\sqrt{\vec{k}^2+a_t^2m^2-a_t^2R/6}
\label{Tomega}
\end{equation}
such that as we approach the asymptotically dS spacetime $\widetilde{\mathcal{M}_{\Sigma}}$
\begin{equation}
\lim_{\eta\rightarrow-\infty}\mbox{ }T_{\vec{k}}(\eta)=\frac{1}{\sqrt{2k}}\mbox{ exp}(-ik\eta)
\label{bdc}
\end{equation}
gives the Bunch--Davies vacuum. This is consistent with a bulk--to--boundary correspondence via the injective $*-$homomorphism $\alpha_f\mbox{ : }\mathcal{A}(\widetilde{\mathcal{M}_{\Sigma}})\rightarrow\mathcal{A}({\mathfrak{J}^{-}})$ in order to construct an induced Hadamard ground state, i.e. Bunch--Davies, on the bulk FRW spacetime \cite{Dap09a,Dap09b}.

\subsection{Excited States as Generalized Hadamard States\label{ESGHS}}

The formulation of a generalized free field state in the algebraic framework, 
\begin{equation}
\omega^B(A_xA_{y}):= \frac{\omega(B_xA_xA_{y}B_{y})}{\omega(B_xB_{y})}
\end{equation}
follows from a generalized Hadamard condition such that any finite excitation of a free field Hadamard state is itself a Hadamard state \cite{San10}. For example, in Minkowski spacetime the free field Kubo--Martin--Schwinger (KMS) state is indeed Hadamard and invariant under the $*-$automorphisms $\alpha_t$ such that, given the global temperature parameter $\beta^{-1}$,
\begin{equation}
\omega(\mbox{ }\alpha_t(A(f))A(g)\mbox{ })=\omega(\mbox{ }A(g)\alpha_{t-i\beta}(A(f))\mbox{ })
\end{equation}
where
\begin{equation}
\alpha_t(A(f)) := A(\mbox{ }f(\tau_0^{-1}(x_{\mu}))\mbox{ }) 
\end{equation}
for $\tau_0: x_{\mu}\mapsto x_{\mu} +t_{\mbox{ }}\vec{e}_0$ with $\vec{e}_0$ a timelike unit vector. We direct the reader to Refs. \cite{Lin13,Fre14} for a rigorous and extensive treatment of both the vacuum and the thermal KMS state, constructed at a finite time in a Hamiltonian approach to perturbative algebraic quantum field theory in Minkowski spacetime via a distinguished time-direction using a one-parameter group of automorphisms $\alpha_t$, where the interacting dynamics are related to free dynamics by a co-cycle in the algebra of the free field; for a similar treatment of non-equilibrium steady states see Ref. \cite{Hac18}. However, in FRW spacetimes there is no time translation invariance and hence no abelian one-parameter group of automorphisms $\alpha_t$ implemented as unitary operators on a corresponding Fock space \cite{Haa84}, i.e. there is no well defined Hamiltonian as the generator of time translations and no strict notion of local thermal equilibrium in non-stationary spacetimes. This has led to several innovative and interesting frameworks, e.g. the Almost Equilibrium States of Ref. \cite{Kus08}, Local $S_x$ Thermal Equilibrium States found in Refs. \cite{Buc02S,Sch08}, and the Bulk-to-Boundary Approximate KMS States in Ref. \cite{Dap11}.  

In this work, we invoke the notion of a propagator-family \cite{Oji86,Buc02} as a non-commutative two-parameter family of automorphisms $\alpha_{t,s}$ such that $\alpha_{t,r} = \alpha_{t,s}\circ\alpha_{s,r}$ and the following group automorphism properties are imposed to ensure the dynamics are consistent with a causal propagator: 
\begin{eqnarray}
\alpha_{t,t} &=& \mathbbm{1}\\
\alpha_{t,s}^{-1} &=& \alpha_{s,t}\\
\alpha_{t,s}(A_tB_t) &=& \alpha_{t,s}(A_t)\mbox{ }\alpha_{t,s}(B_t).
\end{eqnarray}
We define the evolution of the state via the composition
\begin{equation}
\omega_t(A) := \omega_s(A)\circ\beta_{s,t} = \omega(\mbox{ }\alpha_{t,s}(A_s)\mbox{ })
\end{equation}
where $\beta_{r,t} = \beta_{r,s}\circ\beta_{s,t}$. The infinitesimal generators of time shifts are then defined via the relations
\begin{eqnarray}
\dot{\alpha}_{t,s} &=& d_t\circ \alpha_{t,s}\\
\dot{\beta}_{s,t} &=& \beta_{s,t}\circ\delta_t
\end{eqnarray}
where 
\begin{eqnarray}
d_t :&=& \lim_{\Delta t\rightarrow0}\frac{\alpha_{t+\Delta t,t}-\alpha_{t,t}}{\Delta t}\label{da}\\
\delta_t :&=& \lim_{\Delta t\rightarrow0}\frac{\beta_{t,t+\Delta t}-\beta_{t,t}}{\Delta t}\label{db}
\end{eqnarray}
such that 
\begin{equation}
\dot{\alpha}_{t,s}(A_sB_s) = \dot{\alpha}_{t,s}(A_s)B_t + A_{t\mbox{ }}\dot{\alpha}_{t,s}(B_s).
\end{equation}
We may not equate Eq. (\ref{da}) with the Heisenberg equation of motion in non-stationary spacetimes; however, we may define a generator of a perturbed time shift via the relation 
\begin{equation}
\delta_{t}^P(A) := [iP_t\mbox{, }A]
\end{equation}
given
\begin{equation}
\dot{\beta}_{s,t}^P = \beta_{s,t}^P\circ(\delta_t+\delta_t^P).
\end{equation}
with $\beta_{t,t}^P=\mathbbm{1}$ and time dependent perturbation $P_t$. Hence, we let
\begin{equation}
\beta_{t_i,t_f}^P(A_{t_i}) := \mathfrak{U}(t_f,t_i)^{-1}\beta_{t_i,t_f}(A_{t_i})\mbox{ }\mathfrak{U}(t_f,t_i)
\end{equation}
where
\begin{equation}
\mathfrak{U}(t_f,t_i):=\mbox{T}\bigg[\mbox{exp}\bigg(-i\int_{t_i}^{t_f}dt\mbox{ }\beta_{s,t}(P_s)\bigg)\bigg]
\end{equation}
with $\mbox{T}[...]$ as the time ordered product and $\mathfrak{U}(t_f,t_i)^{-1}:=\mathfrak{U}(t_i,t_f)$ such that 
\begin{equation}
\omega_{t_f}^P(A):= \omega(\mbox{ }\beta_{t_i,t_f}^P(A_{t_i})\mbox{ }) = \omega(\mbox{ }\alpha_{t_f,t_i}(A_{t_i})\mbox{ })\circ\gamma_{t_i,t_f}
\end{equation}
given $\gamma_{t_i,t_f}:= \mbox{Ad }\mathfrak{U}(t_f,t_i)^{-1}$. The generalized excited state may now be written as
\begin{equation}
\omega_{t_f}^P(AA) = \frac{ \omega(\mbox{ }\beta_{t_f,t_i}^P(A_{t_i}A_{t_i})\mbox{ })}{ \omega(\mbox{ }\beta_{t_f,t_i}^P(\mathbbm{1})\mbox{ })} = \frac{ \omega(\mbox{ }\alpha_{t_f,t_i}(A_{t_i})\mbox{ }\alpha_{t_f,t_i}(A_{t_i})\mbox{ })\circ\gamma_{t_i,t_f}} { \omega(\mathbbm{1})\circ\gamma_{t_i,t_f}}.\label{pstate1}
\end{equation}

\subsection{Excited States via Generalized Perturbative Interactions\label{GESPI}}

We begin with the classical Lagrangian $\mathscr{L}=\mathscr{L}_0+\mathscr{L}_I$ given an interaction term of the general form 
\begin{equation}
\mathscr{L}_I:=-\sum_i\kappa_i\Phi_i,
\end{equation}
where $\kappa_i$ as a perturbative coupling parameter and $\Phi_i$ as any polynomial in the field $\phi$. Interacting time ordered products as elements of the free field algebra are in general expressed via Bogoliubov's formula
\begin{equation}
\mbox{T}\bigg[\prod_{i=1}^m\int d\mu_i\mbox{ }f_i\Phi_i\bigg] = \sum_n\frac{i^n}{n!}\mbox{R}_n\bigg[\prod_{i=1}^m\int d\mu_i\mbox{ }f_i\Phi_i\mbox{; }\int d\mu\mbox{ }\theta\mathscr{L}_I^{\otimes n}\bigg],
\end{equation}
where $\mbox{R}_n[...]$ is the retarded product to order $n$, as defined in Sec. 4.1 of ref \cite{Hol05} and $\theta\in\mathcal{D}(\mathcal{M}_{\mbox{\bf{g}}})$ a smooth function of compact support. This is of course a well studied perturbative power series with no expectation of convergence and we do not rigorously prove the existence of $P_t$ here. Instead, we invoke the axioms and analysis of Ref. \cite{Hol05} such that if the perturbative quantum field theory satisfies the field equations in the presence of an arbitrary classical current source $J(x_{\mu})$ we may at least rely on Wick polynomials $W^J\in\mathcal{W}(\mathcal{M}_{\Sigma}, \mbox{\textbf{g}},J)$ as self-interactions in the form of an arbitrary but finite $n^{th}$ order perturbative correction to $\phi$ and more generally on the existence of time ordered products and a conserved stress-energy tensor. Hence, we follow Ref. \cite{Hol05} in constructing an interacting theory with $\mathscr{L}_I$ given by the very general, yet nontrivial, classical interaction Lagrangian  
\begin{equation}
\mathscr{L}_I = -J(x_{\mu})\phi(x_{\mu}).
\end{equation}
The interacting quantum theory, now generated by elements of $\mathcal{W}(\mathcal{M}_{\Sigma}, \mbox{\textbf{g}},J)$, is constructed such that eqs. (\ref{bua1}) and (\ref{bua2}) remain satisfied by the sourced $A^J(f)$ and $\mathcal{W}(\mathcal{M}_{\Sigma}, \mbox{\textbf{g}},J)\rightarrow\mathcal{W}(\mathcal{M}_{\Sigma}, \mbox{\textbf{g}})$ via the relation
\begin{equation}
A^J(\widehat{K}f) = \int d\mu_{\mbox{\textbf{g}}}\mbox{ }f(x_{\mu})J(x_{\mu})\cdot\mathbbm{1}
\end{equation}
where Eq. (\ref{bua3}) is recovered in the case of a vanishing source. 

\begin{figure}
\begin{center}
\includegraphics[scale=0.3]{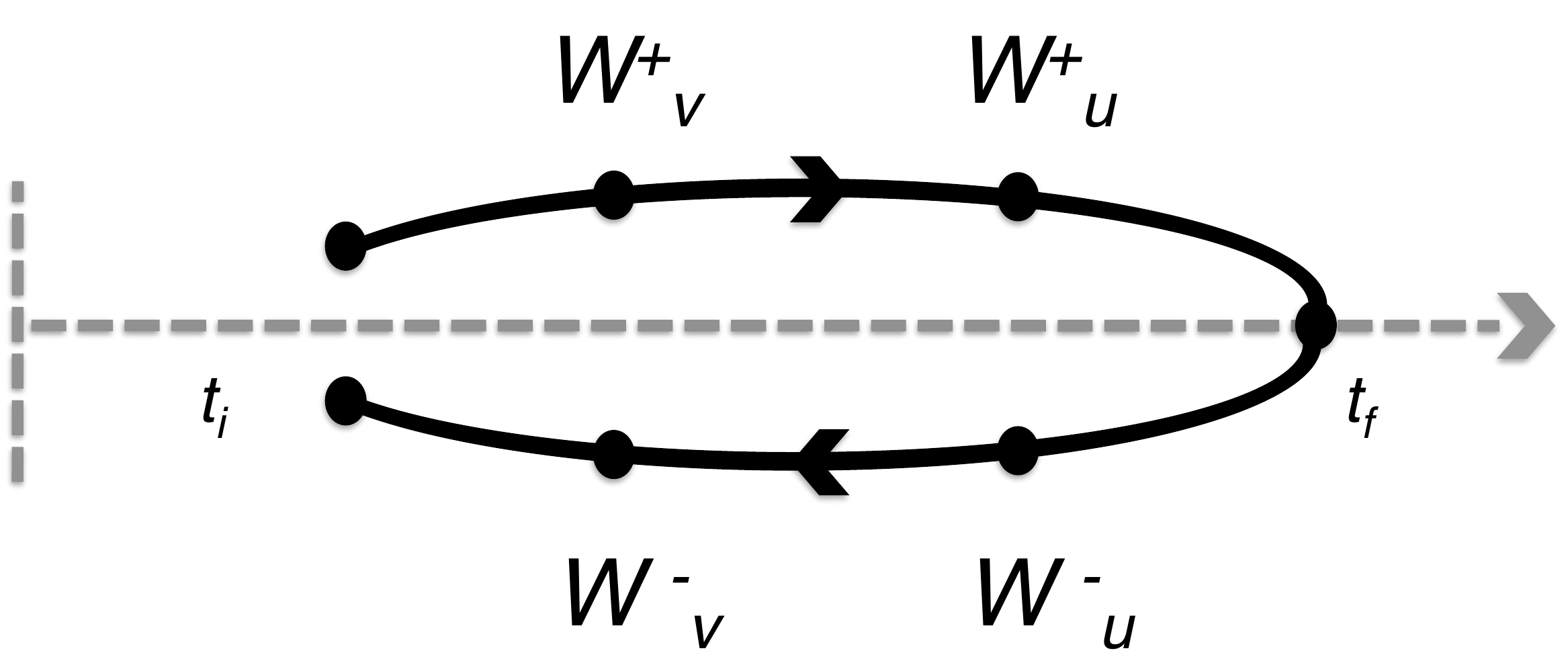}
\caption{\footnotesize{Closed--time--path evolution for the finite macroscopic cosmological time interval $t_i,t_f\in I_t$, given $t_i<t_v<t_u<t_f$ on the forward(+) branch $t_f<t_u<t_v<t_i$ on the backward(-) branch.}}
\label{CTP0}
\end{center}
\end{figure}

In order to derive a general expression for the excited state $\omega_{t_f}^P(A^J(f)A^J(g))$ we express the time averaged perturbation as
\begin{equation}
P_{t_f} = \kappa \int_{t_i}^{t_f}dt_u\int d^3u\mbox{  }\theta(t_u,\vec{u})W_u^J
\end{equation} 
with $\theta(t_u,\vec{u}) = h(t_u)\psi(\vec{u})$ such that the adiabatic limit corresponds to the constant function
\begin{equation}
\psi(\vec{u}) = 1\mbox{ on supp } f\subset\mathcal{M}_{\Sigma}^{(I_t)} = \{(t_u,\vec{u})|\mbox{ }t_i < t_u < t_f\}.  
\end{equation}
Hence,
\begin{eqnarray}
\mathfrak{U}(t_f,t_i)&=&\mathbb{1}-i\kappa\int_{t_i}^{t_f}dt_u\int d^3u\mbox{  }h(t_u)W_u^J\\\nonumber
& &-\mbox{ }\frac{\kappa^2}{2}\int_{t_i}^{t_f}dt_u\int_{t_i}^{t_f}dt_v\int d^3u\int d^3v\mbox{  }h(t_u)h(t_v)\mbox{ T}\bigg[W_u^JW_v^J\bigg]
\end{eqnarray}\\
truncated to second order in $\kappa$. The perturbed state $\omega_{t_f}^P(AA)$ of Eq. (\ref{pstate1}), rewritten as  
\begin{eqnarray}
\omega_{t_f}^P(A^J(f)A^J(g)) &=& \frac{ \omega\bigg(\mbox{ }\mathfrak{U}^{-1}(t_i,t_f)\mbox{ }A^J(f)^JA(g)\mbox{ }\mathfrak{U}(t_i,t_f)\mbox{ }\bigg) }{ \omega\bigg(\mbox{ }\mathfrak{U}^{-1}(t_i,t_f)\mbox{ }\mathbb{1}\mbox{ }\mathfrak{U}(t_i,t_f)\mbox{ }\bigg) },\nonumber\\
& &
\end{eqnarray}
may now be expressed
\begin{eqnarray}
\omega_{t_f}^P(A^J(f)A^J(g)) &=& \Bigg\{\mbox{ }\omega(A^J(f)A^J(g))+\omega\bigg(\frac{\kappa^2}{2}\int_{t_i}^{t_f}dt_u\int d^3u\int_{t_i}^{t_f}dt_v\int d^3v\mbox{ }h(t_u)h(t_v)\bigg[\nonumber\\
& & \mbox{ }\mbox{ }\mbox{ }\overline{\mbox{T}}[W_u^J]\mbox{ }A^J(f)\mbox{ }A^J(g)\mbox{ T}[W_v^J] - A^J(f)\mbox{ }A^J(g)\mbox{ T}[W_u^J\mbox{ }W_v^J]\nonumber\\
& & +\mbox{ }\mbox{T}[W_u^J]\mbox{ }A^J(f)\mbox{ }A^J(g)\mbox{ }\overline{\mbox{T}}[W_v^J] - \overline{\mbox{T}}[W_u^J\mbox{ }W_v^J]\mbox{ }A^J(f)\mbox{ }A^J(g)\mbox{ }\bigg] \bigg)\mbox{ }\Bigg\}\nonumber \\
&\times&\mbox{ }\Bigg\{\mbox{ }\omega(\mathbbm{1}) + \omega\bigg(\frac{\kappa^2}{2}\int_{t_i}^{t_f}dt_u\int d^3u\int_{t_i}^{t_f}dt_v\int d^3v\mbox{ }h(t_u)h(t_v)\bigg[\nonumber\\
& & \mbox{ }\mbox{ }\mbox{ }\overline{\mbox{T}}[W_u^J]\mbox{ }\mbox{T}[W_v^J] - \mbox{T}[W_u^J\mbox{ }W_v^J] +\mbox{T}[W_u^J]\mbox{ }\overline{\mbox{T}}[W_v^J] - \overline{\mbox{T}}[W_u^J\mbox{ }W_v^J]\mbox{ }\bigg] \bigg)\Bigg\}^{-1}\nonumber\\
& &
\label{geqs2}
\end{eqnarray}
or
\begin{eqnarray}
{\omega}_{t_f}^P(A(f)A(g)) &=& Z_{\omega}\bigg\{\omega_{t_f}(A(f)A(g))\nonumber\\
&+&\omega\bigg( \frac{\kappa^2}{2}\int_{t_i}^{t_f}dt_x\int d^3x\int_{t_i}^{t_f}dt_u\int d^3u\int_{t_i}^{t_f}dt_v\int d^3v\int_{t_i}^{t_f}dt_y\int d^3y\bigg[\nonumber\\
& &\mbox{ }f(t_x,\vec{x})h(t_u)h(t_v)g(t_y,\vec{y})\bigg(\nonumber\\
& &W_{u}^{-}A_x^{-}A_y^{+}W_{v}^{+} - A_x^{-}A_y^{+}W_{u}^{+}W_{v}^{+} + W_{v}^{-}A_x^{-}A_y^{+}W_{u}^{+} - W_{v}^{-}W_{u}^{-}A_x^{-}A_y^{+}\bigg)\bigg)\bigg]\bigg\}\nonumber\\
& &
\label{omgP}
\end{eqnarray}
where
\begin{eqnarray}
Z_{\omega} := 1 &-& \omega_{t_f}\bigg(\frac{\kappa^2}{2}\int_{t_i}^{t_f}dt_u\int d^3u\int_{t_i}^{t_f}dt_v\int d^3v\mbox{ }h(t_u)g(t_v)\bigg[\nonumber\\
& &\mbox{ }\mbox{ }\mbox{ }\mbox{ }\mbox{ }\mbox{ }\mbox{ }\mbox{ }\mbox{ }\mbox{ }W_{u}^-W_{v}^+ - W_{u}^+W_{v}^+ + W_{v}^-W_{u}^+ - W_{v}^-W_{u}^-\bigg]\bigg)
\end{eqnarray}
is a state dependent normalization factor. Here, we replace the superscript $J$ of the sourced field with the time-ordering index $\pm$ corresponding to the forward(+) and backward(-) branch of the closed-time-path depicted in Fig. \ref{CTP0}. In addition, we relabel $A_x^J$ as $A_{x}^{-}$ and $A_{y}^J$ as $A_{y}^{+}$ to denote that no point $x_{\mu}\in\mbox{supp }f(x_{\mu})$ is in the past of $y_{\mu}\in\mbox{supp }g(y_{\mu})$. This is equivalent to the Schwinger--Keldysh ``in--in'' formalism and hence appropriate for far-from-equilibrium interactions. Notably, the CPT theorem in FRW spacetimes relates an in-state in an expanding universe to an in-state in the corresponding contracting universe \cite{Hol10}. We also note that both the excitation and the perturbative portion of the normalization factor, i.e. the terms proportional to $\kappa^2$, are finite via the properties of Hadamard states. 

Equivalently, we may write the symmetric part of the perturbed state as a homogeneous and isotropic quasifree state restricted to the Cauchy surface $\Sigma_{t}$
\begin{eqnarray}
\omega_{t}^{S,P}(A(f)A(g)) &=& 8\pi \int dk\mbox{ }k^2\mbox{ }\mathfrak{X}_{\vec{k}}^{P_t}\mbox{ }|X_{\vec{k}}(t)|^2\mbox{ }\overline{ \hat{\overline{f}}_{\vec{k}}}\mbox{ }\hat{g}_{\vec{k}}
\label{pstate2}
\end{eqnarray}
for all $t > t_f$ where, for example, the source $J(x_{\mu})$ vanishes for $t\notin I_t$. Here, $\mathfrak{X}_{\vec{k}}^{P_t}$ is now the polynomialy bounded function perturbed via the generator $P_t$ and $X_{\vec{k}}(t)$ the mode functions defined in Eq. (\ref{bt1}), such that
\begin{eqnarray}
\mathfrak{X}_{\vec{k}}^{P_t} &=& Z_{\omega}\bigg[\mathfrak{X}_{\vec{k}}^0 + \kappa^2\mbox{ }\omega_{t}\bigg(\widehat{\mathfrak{D}}_{k}\bigg[\int_{t_i}^{t}dt_x\int d^3x\int_{t_i}^{t_f}dt_u\int d^3u\int_{t_i}^{t_f}dt_v\int d^3v\int_{t_i}^{t}dt_y\int d^3y\bigg\{\nonumber\\
& &\mbox{ }\mbox{ }\mbox{ }\mbox{ }\mbox{ }f(t_x,\vec{x})g(t_y,\vec{y})h(t_u)h(t_v)\nonumber\\
& &\times\bigg[W_{u}^{-}A_{x}^{-}A_{y}^{+}W_{v}^{+} -  A_{x}^{-}A_{y}^{+}W_{u}^{+}W_{v}^{+} + W_{v}^{-}A_{x}^{-}A_{y}^{+}W_{u}^{+} - W_{v}^{-}W_{u}^{-}A_{x}^{-}A_{y}^{+}\bigg]\bigg\}\bigg)\bigg]\nonumber\\
& &
\label{frakxp}
\end{eqnarray}
given $\mathfrak{X}_{\vec{k}}^0$ as the polynomially bounded function, as defined in Eq. (\ref{hiqs0}), for the state $\omega_{0}(A(f)A(g))$ specified at the time $t_i$ and the differential operator $\widehat{\mathfrak{D}}_{k}$ defined as
\begin{equation}
\widehat{\mathfrak{D}}_{k} := \bigg(8\pi\mbox{ } k^2\mbox{ }|X_{\vec{k}}(t)|^2\mbox{ }\overline{ \hat{\overline{f}}_{\vec{k}}}\mbox{ }\hat{g}_{\vec{k}}\bigg)^{-1}\frac{d}{dk}.
\end{equation}

\section{Renormalized Energy Density from the Algebraic State\label{REDAS}}

Following the formulations and results in Refs. \cite{Sch08,Deg13,Hac15}, we now review the expectation value of the energy density of a minimally coupled, i.e. $\xi=0$, free scalar field in an arbitrary mixed state propagating in a non-stationary FRW spacetime background. We begin with the expectation value of the renormalized stress-energy tensor, taken in the free field limit at a finite a cosmological time $t>t_f$ and restricted to the total diagonal such that
\begin{eqnarray}
\omega(\mbox{ T}_{\mu\nu}(:A_{x}^2:)\mbox{ })&=&\bigg\{\omega^{S,P}\bigg(\widehat{D}_{x,y}[A_{x}A_{y}]\bigg) - \widehat{D}_{x,y}\mathbbm{H}_1^s(x_{\mu},y_{\mu})+\mbox{ }\frac{1}{3}\widehat{K}_{x}\mbox{ }\mathbbm{H}_1^s(x_{\mu},y_{\mu})+C_{\mu\nu}(x_{\mu})\bigg\}\bigg|\begin{matrix} & &\\x_{\mu}=y_{\mu}\end{matrix}.\nonumber\\
\label{ped}
\end{eqnarray}
Here, the bi--differential operator $\widehat{D}$ is defined
\begin{equation}
\widehat{D}_{a,b} := \frac{1}{2}\bigg(\frac{\partial}{\partial_{0}^a}\frac{\partial}{\partial_{0}^b} +\frac{1}{a_t^2}\nabla^a\nabla^b + m^2\bigg)
\end{equation}
and the purely geometric Hadamard parametrix is expressed
\begin{equation}
\mathbbm{H}_n(x_{\mu},y_{\mu}) = \lim_{\epsilon\downarrow0}\frac{1}{4\pi^2}\bigg[\mbox{ }\frac{1}{\sigma_{\epsilon}(x_{\mu},y_{\mu})}+\frac{1}{L^2}\sum_{m=1}^{n}V_m\bigg(\frac{\sigma(x_{\mu},y_{\mu})}{L^2}\bigg)^m\mbox{ log}\bigg(\frac{\sigma_{\epsilon}(x_{\mu},y_{\mu})}{L^2}\bigg)\mbox{ }\bigg],
\end{equation}
where $V_m$ satisfies the so called Hadamard recursion relations (see e.g. Ref. \cite{Sch08}).
\begin{equation}
\mathbbm{H}_1^s(x_{\mu},y_{\mu}) := \frac{1}{2}\bigg(\mathbbm{H}_1(x_{\mu},y_{\mu}) + \mathbbm{H}_1(y_{\mu},x_{\mu})\bigg) 
\end{equation}
is then the symmetric Hadamard bi--distribution truncated to order $n=1$ where 
\begin{equation}
V_1=-\frac{1}{3}\widehat{K}_x\mathbbm{H}_1^s(x_{\mu},y_{\mu})
\end{equation}
and $C_{\mu\nu}(x_{\mu})$ carries the renormalization freedom of Wick products contained in a conserved stress-energy tensor.

\subsection{General Form from the Perturbed Stated\label{TGF}}

The renormalized energy density of the perturbed state, taken in the free field limit for $\eta(t)>\eta_f$, is found via the restriction of the stress-energy tensor; first to the partial diagonal $\eta_x=\eta_y=\eta(t)$ then in the coincidence limit $\vec{x}=\vec{y}$ such that
\begin{equation}
\mathscr{E}_A^{P_{\eta(t)}} := \omega_{\eta(t)}^P(\mbox{ T}_{00}(:A^2(f):)\mbox{ }).
\end{equation}
This is a nuanced expression that we briefly explain term by term. Here,
\begin{eqnarray}
\omega^{P}\bigg(\widehat{D}_{x,y}[A(f)A(g)]\bigg)\bigg|_{x=y} &=&\frac{1}{2\pi^2}\int_0^{\infty}dk\mbox{ }\bigg\{k^2\mbox{ }\mathfrak{X}_{\vec{k}}^{P_{\eta(t)}}\mbox{ }\nonumber\\
&\times& \frac{1}{a_t^4}\bigg[|X_{\vec{k}}'(\eta)|^2 - a_tH_t(|X_{\vec{k}}(\eta)|^2)'+\mbox{ }(k^2+a_t^2m^2+a_t^2H_t^2)|X_{\vec{k}}(\eta)|^2\bigg]\bigg\}\nonumber\\
 & &
\label{dmint}
\end{eqnarray}
is the divergent mode integral with mode functions $X_{\vec{k}}(\eta)$ found via SLE minimized energy density of the ground state and the polynomially bounded function $\mathfrak{X}_{\vec{k}}^{P_\eta(t)}$ determined by the perturbation $P_{\eta(t)}$ via Eq. (\ref{frakxp}) of the previous subsection.
\begin{eqnarray}
\widehat{D}_{x,y}\mathbbm{H}_1^s(f,g)|_{x=y} &=& \frac{1}{4\pi^2}\bigg[-\frac{1}{a_t^4}\frac{2}{r_+^4}(f)+\frac{m^2+H_t^2}{2a_t^2}\frac{1}{r_+^2}(f)\nonumber\\
& &\mbox{ }+\mbox{ }\bigg(\frac{m^4}{16}-\frac{2m^2H_t^2}{16}+\frac{2\ddot{H}_tH_t}{16}+\frac{6\dot{H}_tH_t^2}{16}-\frac{\dot{H}_t^2}{16}\bigg)\bigg(\mbox{lo}_0(f)+\mbox{log}(a_t^2)\bigg)\nonumber\\
& &\mbox{ }+\mbox{ }\frac{\square_{\mbox{g}}R}{120}+m^2\bigg(\frac{7H_t^2}{24}+\frac{\dot{H}_t}{4}\bigg) - \frac{m^4}{8} + \frac{H_t^4}{80} - \frac{11H_t\ddot{H}_t}{120}\nonumber\\
& &\mbox{ }-\mbox{ }\frac{61H_t^2\dot{H}_t}{120}-\frac{19\dot{H}_t^2}{240}\bigg],
\end{eqnarray}
where the singular counter--terms given by the symmetric distributions $r_+^4$, $r_+^2$, and $\mbox{lo}_0$ are defined via the convolutions
\begin{eqnarray}
\frac{2}{r_+^4}(f) &:=& \lim_{\epsilon\rightarrow+0}\int_{\mathbbm{R}^3}d^3x\mbox{ }\frac{\nabla f(\vec{x})}{\vec{x}^2+\epsilon^2}\\
\frac{1}{r_+^2}(f) &:=& \lim_{\epsilon\rightarrow+0}\int_{\mathbbm{R}^3}d^3x\mbox{ }\frac{f(\vec{x})}{\vec{x}^2+\epsilon^2}\\
\mbox{lo}_0(f) &:=& \int_{\mathbbm{R}^3}d^3x\mbox{ }f(\vec{x})\mbox{ log}(\vec{x}^2),
\end{eqnarray}
for a fixed $f\in C_0^{\infty}(\mathbbm{R}^3)$ are the geometric contribution of the parametrix. Here, the sum of the singular terms may then be rewritten as a mode integral, i.e. 
\begin{eqnarray}
&\lim_{\epsilon\rightarrow0}&\frac{1}{2\pi^2}\int dk\mbox{ }k^2I(k)\mbox{ exp}(i\vec{k}\cdot\vec{x})\mbox{ exp}(-k\epsilon)=\nonumber\\
& & \frac{1}{2\pi^2}\bigg\{-\mathfrak{C}_{-1}\frac{2}{r_+^4}(f) + \mathfrak{C}_0\frac{1}{r_+^2}(f) + \mathfrak{C}_1\mbox{lo}_0(f)\nonumber\\
&&+4\pi\int_{\mathbbm{R}^3} d^3x\mbox{ }f(\vec{x})\mbox{ }\lim_{M\rightarrow\infty}\bigg[\int_0^M dk\mbox{ }k\bigg(k\mbox{ }I(k)\nonumber\\
& & -\mathfrak{C}_{-1}k^2-\mathfrak{C}_{0}\bigg) - \mathfrak{C}_1\bigg(\mbox{ log}(ML) - 1 +\gamma_{EM} \bigg)\mbox{ }\bigg]\mbox{ }\bigg\}\nonumber\\
& &
\label{hpmi}
\end{eqnarray}
where $\gamma_{EM}$ is the Euler--Mascheroni constant and the integrand $I(k)$ has asymptotic behavior
\begin{equation}
I(k\rightarrow\infty) = \sum_{m=-1}^1\frac{\mathfrak{C}_m}{k^{2m+1}} + \mathcal{O}(k^{-5}),
\end{equation}
such that the subtraction of singular terms may occur inside the mode integral of Eq. (\ref{dmint}); and
\begin{eqnarray}
\frac{1}{3}\widehat{K}_{\eta}\mathbbm{H}_1^s(f,g)|_{x=y} &=& \frac{1}{4\pi^2}\bigg(\frac{3\dot{H}_t^2}{40} +\frac{\dddot{H}_t}{20} + \frac{7H_t^2\dot{H}_t}{60} + \frac{7H_t\ddot{H}_t}{20}\nonumber\\
& & - \frac{29H_t^4}{60} - \frac{m^4}{8} + \frac{m^2H_t^2}{2} + \frac{m^2\dot{H}_t}{4}\bigg).
\end{eqnarray}
\begin{eqnarray}
C_{00}(\mbox{ }\eta(t)\mbox{ }) &:=& \mathfrak{c}_1m^4\mbox{g}_{00} +\mathfrak{c}_2m^2\mbox{G}_{00} +\mbox{ }(3\mathfrak{c}_3+\mathfrak{c}_4)(6\dot{H}_t^2 -12\ddot{H}_tH_t-36\dot{H}_tH_t^2)\nonumber\\
\end{eqnarray}
allows for a renormalization freedom via the coefficients $\mathfrak{c}_{\{1,2,3,4\}}$, which are not fixed \emph{a priori} in the theory. However, they may be constrained either by experiment or physical arguments. This is to say that $\mathfrak{c}_1$ and $\mathfrak{c}_2 $ correspond to a renormalization of the cosmological constant and Newton's constant respectively, as quantities appearing in Einstein's equation, while the sum $(3\mathfrak{c}_3+\mathfrak{c}_4)$ is constrained by higher order derivative corrections to the semiclassical approximation. In this work we take the position that $\mathfrak{c}_{\{2,3,4\}}$ are not free parameters at the length scale, $L$ of Eq. (\ref{hadpar}), probed by current experiments that support the $\Lambda$CDM model and we omit the afforded freedom. However, we do embrace renormalization of the vacuum energy density where the requirement that this scheme reduces to normal ordering \cite{Hol05,Hol08b,Deg13}, i.e. subtraction of $\mathscr{E}_{A}^0$ as the reference state in Minkowski spacetime 
\begin{equation}
\mathscr{E}_{A,t_i}^0 = \frac{1}{2}\int_0^{\infty}\frac{d^3k}{(2\pi)^3}\mbox{ }\Bigg\{|T_{\vec{k}}(t_i)'|^2 + \Omega_{\vec{k}}|T_{\vec{k}}(t_i)|^2\Bigg\}
\label{refsta}
\end{equation}
fixes $\mathfrak{c}_1$ as a function of $L$ such that
\begin{equation}
\mathfrak{c}_1(L)\mbox{ }m^4\mbox{g}_{00} = -\frac{m^4}{32\pi^2}\bigg(\mbox{log}(mL)-\mbox{log}(2)-\frac{3}{4}+\gamma_{EM}\bigg)\mbox{g}_{00}.
\label{der}
\end{equation}

Hence, we find as our main result the general expression for $\mathscr{E}_{A}^{P_{\eta(t)}}$ as the renormalized, perturbed energy density of a massive, minimally coupled scalar field in the free field limit to be
\begin{eqnarray}
\mathscr{E}_{A}^{P_{\eta(t)}} &=& \frac{1}{2\pi^2}\int_0^{\infty}dk\bigg\{\mbox{ }k^2\mbox{ }\frac{1}{a_t^4}\bigg[|X_{\vec{k}}'(\eta)|^2 - a_tH_t(|X_{\vec{k}}(\eta)|^2)'+\mbox{ }(k^2+a_t^2m^2+a_t^2H_t^2)|X_{\vec{k}}(\eta)|^2\bigg]\nonumber\\
&\times&\mbox{ }Z_{\omega}\bigg(\mathfrak{X}_{\vec{k}}^{0} + \mbox{ } \omega_{\eta(t)}\bigg(\int_{\eta_i}^{\eta(t)}d\eta_x\int d^3x\int_{\eta_i}^{\eta_f}d\eta_u\int d^3u\int_{\eta_i}^{\eta_f}d\eta_v\int d^3v\int_{\eta_i}^{\eta(t)}d\eta_y\int d^3y\bigg\{\nonumber\\
& &\mbox{ }\mbox{ }\mbox{ }\mbox{ }f(\eta_x,\vec{x})g(\eta_y,\vec{y})h(\eta_u)h(\eta_v)\nonumber\\
& &\mbox{ }\mbox{ }\mbox{ }\mbox{ }\times\mbox{ }\widehat{\mathfrak{D}}_k\bigg[W_{u}^{-}A_{x}^{-}A_{y}^{+}W_{v}^{+} -  A_{x}^{-}A_{y}^{+}W_{u}^{+}W_{v}^{+} + W_{v}^{-}A_{x}^{-}A_{y}^{+}W_{u}^{+} - W_{v}^{-}W_{u}^{-}A_{x}^{-}A_{y}^{+}\bigg]\bigg)\bigg|_{x=y}\bigg\}\bigg)\nonumber\\
&-&k\frac{1}{2a_t^4}-\frac{1}{k}\frac{H_t^2+m^2}{4a_t^2}-\frac{1}{k^3}\bigg(\frac{m^4-2m^2H_t^2+2\ddot{H}_tH_t+6\dot{H}_tH_t^2-\dot{H}_t^2}{16}\bigg)\bigg\}\nonumber\\
&-&\frac{m^2H_t^2}{96\pi^2} - \frac{m^4(1-4\mbox{log}(2))}{128\pi^2} + \frac{12H_t^4 + 48H_t^2\dot{H}_t + 36\dot{H}_t^2}{96\pi^2}.\nonumber\\
& &
\label{mainres}
\end{eqnarray}

\subsection{Cubic Interaction Example\label{CIE}}

As an example of a concrete realization of Eq. (\ref{mainres}) that is in principal amenable to a numerical calculation we choose here the perturbation $P_t$ to be the product 
\begin{equation}
P_{\eta} = \kappa\int_{\eta_i}^{\eta_f}d\eta_u\int d^3u\mbox{ }\theta(\eta_u,\vec{u})A_uB_uC_c
\end{equation}
corresponding to a trilinear scalar interaction with classical Lagrangian
\begin{equation}
\mathscr{L}_I=-\kappa\phi_1\phi_2\phi_3.
\end{equation}
We concede that such a product is not the self-interacting Wick polynomial $W_u$ employed in the previous section, however we believe this example highlights key features of the perturbed energy density and is thus useful. In this instance, the perturbed stated $\omega_{\eta(t)}^P(A(f)A(g))$ may be written
\begin{eqnarray}
\omega_{\eta(t)}^P(A(f)A(g)) &=& Z_{\omega}\bigg\{\omega_{\eta(t)}^0(A(f)A(g))\nonumber\\
&+&\omega\bigg(\frac{\kappa^2}{2}\int_{\eta_i}^{\eta(t)}d\eta_x\int d^3x\int_{\eta_i}^{\eta_f}d\eta_u\int d^3u\int_{\eta_i}^{\eta_f}d\eta_v\int d^3v\int_{\eta_i}^{\eta(t)}d\eta_y\int d^3y\bigg[\nonumber\\
& &f(\eta_x,\vec{x})g(\eta_y,\vec{y})h(\eta_u)\psi(\vec{u})h(\eta_v)\psi(\vec{v})\nonumber\\
& &\times\mbox{ }\bigg(A_{u}^{-}B_{u}^{-}C_{u}^{-}A_{x}^{-}A_{y}^{+}A_{v}^{+}B_{v}^{+}C_{v}^{+} - A_{x}^{-}A_{y}^{+}A_{u}^{+}B_{u}^{+}C_{u}^{+}A_{v}^{+}B_{v}^{+}C_{v}^{+}\nonumber \\
& & + \mbox{ }A_{v}^{-}B_{v}^{-}C_{v}^{-}A_{x}^{-}A_{y}^{+}A_{u}^{+}B_{u}^{+}C_{u}^{+} -\mbox{ }A_{v}^{-}B_{v}^{-}C_{v}^{-}A_{u}^{-}B_{u}^{-}C_{u}^{-}A_{x}^{-}A_{y}^{+}\bigg)\bigg]\bigg)\bigg\}
\label{pstate4}
\end{eqnarray}
where
\begin{eqnarray}
Z_{\omega} = 1 &-& \omega\bigg(\frac{\kappa^2}{2}\int_{\eta_i}^{\eta_f}d\eta_u\int d^3u\int_{\eta_i}^{\eta_f}d\eta_v\int d^3v\int d^3y\bigg\{h(\eta_u)\psi(\vec{u})h(\eta_v)\psi(\vec{v})\nonumber\\
&\times&\bigg[A_{u}^{-}B_{u}^{-}C_{u}^{-}A_{v}^{+}B_{v}^{+}C_{v}^{+} - A_{u}^{+}B_{u}^{+}C_{u}^{+}A_{v}^{+}B_{v}^{+}C_{v}^{+}\nonumber \\
&+&A_{v}^{-}B_{v}^{-}C_{v}^{-}A_{u}^{+}B_{u}^{+}C_{u}^{+} - A_{v}^{-}B_{v}^{-}C_{v}^{-}A_{u}^{-}B_{u}^{-}C_{u}^{-}\bigg]\bigg\}\bigg).
\end{eqnarray}
Taking, as an example, the first term of order $\kappa^2$ in Eq. (\ref{pstate4}), defined as $\omega(u^-x^-y^+v^+)$, we may expand it as a homogeneous and isotropic quasifree state such that
\begin{eqnarray}
\omega(u^-x^-y^+v^+)&:=&Z_{\omega}\omega\bigg(\frac{\kappa^2}{2}\int_{\eta_i}^{\eta(t)}d\eta_x\int d^3x\int_{\eta_i}^{\eta_f}d\eta_u\int d^3u\int_{\eta_i}^{\eta_f}d\eta_v\int d^3v\int_{\eta_i}^{\eta(t)}d\eta_y\int d^3y\bigg[\nonumber\\
& &f(\eta_x,\vec{x})g(\eta_y,\vec{y})h(\eta_u)\psi(\vec{u})h(\eta_v)\psi(\vec{v})\mbox{ }A_{u}^{-}B_{u}^{-}C_{u}^{-}A_{x}^{-}A_{y}^{+}A_{v}^{+}B_{v}^{+}C_{v}^{+}\bigg\}\bigg)\nonumber\\
& &\nonumber\\
&=&Z_{\omega}\mbox{ }\omega\bigg(\frac{\kappa^2}{2}\int_{\eta_i}^{\eta(t)}d\eta_x\int d^3x\int_{\eta_i}^{\eta_f}d\eta_u\int d^3u\int_{\eta_i}^{\eta_f}d\eta_v\int d^3v\int_{\eta_i}^{\eta(t)}d\eta_y\int d^3y\bigg[\nonumber\\
& &f(\eta_x,\vec{x})g(\eta_y,\vec{y})h(\eta_u)\psi(\vec{u})h(\eta_v)\psi(\vec{v})\nonumber\\
&\times&\bigg[\omega(A_{u}^{-}A_x^{-})\omega(A_{y}^{+}A_{v}^{+})\omega(B_{u}^{-}B_{v}^{+})\omega(C_{u}^{-}C_{v}^{+})\nonumber\\
&+&\omega(A_x^{-}A_{v}^{+})\omega(A_u^{-}A_{y}^{+})\omega(B_{u}^{-}B_{v}^{+})\omega(C_{u}^{-}C_{v}^{+})\nonumber\\
&+&\omega(A_x^{-}A_{y}^{+})\omega(A_u^{-}A_{v}^{+})\omega(B_{u}^{-}B_{v}^{+})\omega(C_{u}^{-}C_{v}^{+}) \bigg]\bigg\}
\label{term1}
\end{eqnarray}
given the cluster property $\omega(A_uB_v)=\omega(A_u)\omega(B_v)$ $\forall t\notin I_t$. In addition, we may simplify Eq. (\ref{term1}) via a cancellation of terms of the form
\begin{eqnarray}
& &-\mbox{ }\omega(A(f)A(g))\mbox{ }\frac{\kappa^2}{2}\int_{\eta_i}^{\eta_f} d\eta_u\int d^3u\int_{\eta_i}^{\eta_f}d\eta_v\int d^3v\bigg\{h(\eta_u)\psi(\vec{u})h(\eta_v)\psi(\vec{v})\nonumber\\
& &\times\mbox{ }\bigg[\omega(A_{u}^{-}A_{v}^{+})\omega(B_{u}^{-}B_{v}^{+})\omega(C_{u}^{-}C_{v}^{+})\bigg]\bigg\}\nonumber
\end{eqnarray}
by a similar expansion of $Z_{\omega}$ where we now write
\begin{eqnarray}
\omega(u^-x^-y^+v^+)&=&\frac{\kappa^2}{2}\int_{\eta_i}^{\eta(t)}d\eta_x\int d^3x\int_{\eta_i}^{\eta_f}d\eta_u\int d^3u\int_{\eta_i}^{\eta_f}d\eta_v\int d^3v\int_{\eta_i}^{\eta(t)}d\eta_y\int d^3y\bigg[\nonumber\\
& &f(\eta_x,\vec{x})g(\eta_y,\vec{y})h(\eta_u)\psi(\vec{u})h(\eta_v)\psi(\vec{v})\nonumber\\
& &\times\mbox{ }\bigg[\omega(A_{u}^{-}A_x^{-})\omega(A_{y}^{+}A_{v}^{+})\omega(B_{u}^{-}B_{v}^{+})\omega(C_{u}^{-}C_{v}^{+})\nonumber\\
& &\mbox{ }+\mbox{ }\omega(A_x^{-}A_{v}^{+})\omega(A_u^{-}A_{y}^{+})\omega(B_{u}^{-}B_{v}^{+})\omega(C_{u}^{-}C_{v}^{+})\bigg]\bigg\}.
\label{term2}
\end{eqnarray}

Taking the limit $\psi\rightarrow1$ and carrying out the spatial integrals we find
\begin{eqnarray}
& &\omega(u^-x^-y^+v^+)=\nonumber\\
& &\mbox{ } \kappa^2\int d^3k\int d^3p\int_{\eta_i}^{\eta(t)}d\eta_x\int_{\eta_i}^{\eta(t)}d\eta_y\int_{\eta_i}^{\eta_f}d\eta_u\int_{\eta_i}^{\eta_f}d\eta_v\mbox{ }\overline{\hat{\overline{f}}_{\vec{k}}}(\eta_x)\hat{g}_{\vec{k}}(\eta_y)h(\eta_u)h(\eta_v)\bigg\{\nonumber\\
& &\mbox{ }\mathfrak{X}_{\vec{k}}^{\eta(t)}\bigg(\overline{X_{\vec{k}}(\eta_x)}X_{\vec{k}}(\eta_u)+X_{\vec{k}}(\eta_x)\overline{X_{\vec{k}}(\eta_u)}\bigg)\mbox{ }\mathscr{X}_{\vec{k}}^{-+}\bigg(\overline{X_{\vec{k}}(\eta_v)}X_{\vec{k}}(\eta_y)+X_{\vec{k}}(\eta_v)\overline{X_{\vec{k}}(\eta_y)}\bigg)\nonumber\\
& &\times\mbox{ }\mathscr{Y}_{\vec{p}}^{-+}\bigg(\overline{Y_{\vec{p}}(\eta_u)}Y_{\vec{p}}(\eta_v)+Y_{\vec{p}}(\eta_u)\overline{Y_{\vec{p}}(\eta_v)}\bigg)\mbox{ }\mathscr{Z}_{\vec{k}-\vec{p}}^{-+}\bigg(\overline{Z_{\vec{k}-\vec{p}}(\eta_u)}Z_{\vec{k}-\vec{p}}(\eta_v)+Z_{\vec{k}-\vec{p}}(\eta_u)\overline{Z_{\vec{k}-\vec{p}}(\eta_v)}\bigg)\bigg\},\nonumber\\
\label{term3}
\end{eqnarray}
via the construction of the symmetric homogeneous and isotropic Hadamard states of Eq. (\ref{hiqs0}). Here, we introduce a more compact notation with the expression
\begin{eqnarray}
& &\omega_{\eta(t)}(u^-x^-y^+v^+)=\nonumber\\
& &\mbox{ } \kappa^2\int d^3k\int_{\eta_i}^{\eta(t)}d\eta_x\int_{\eta_i}^{\eta(t)}d\eta_y\int_{\eta_i}^{\eta_f}d\eta_u\int_{\eta_i}^{\eta_f}d\eta_v\mbox{ }\overline{\hat{\overline{f}}_{\vec{k}}}(\eta_x)\hat{g}_{\vec{k}}(\eta_y)h(\eta_u)h(\eta_v)\bigg\{\nonumber\\
& &\mbox{ }\mathfrak{X}_{\vec{k}}^{\eta(t)}\bigg(\overline{X_{\vec{k}}(\eta_x)}X_{\vec{k}}(\eta_u)+X_{\vec{k}}(\eta_x)\overline{X_{\vec{k}}(\eta_u)}\bigg)\mbox{ }\mathscr{X}_{\vec{k}}^{-+}\bigg(\overline{X_{\vec{k}}(\eta_v)}X_{\vec{k}}(\eta_y)+X_{\vec{k}}(\eta_v)\overline{X_{\vec{k}}(\eta_y)}\bigg)G_{\vec{k}}^{-+}(\eta_u,\eta_v)\bigg\}\nonumber\\
\label{term4}
\end{eqnarray}
where
\begin{eqnarray}
G_{\vec{k}}^{-+}(\eta_u,\eta_v) &:=& \int d^3p\mbox{ }\bigg\{\mathscr{Y}_{\vec{p}}^{-+}\bigg(\overline{Y_{\vec{p}}(\eta_u)}Y_{\vec{p}}(\eta_v)+Y_{\vec{p}}(\eta_u)\overline{Y_{\vec{p}}(\eta_v)}\bigg)\nonumber\\
& &\mbox{ }\mbox{ }\mbox{ }\mbox{ }\mbox{ }\mbox{ }\times\mbox{ }\mathscr{Z}_{\vec{k}-\vec{p}}^{-+}\bigg(\overline{Z_{\vec{k}-\vec{p}}(\eta_u)}Z_{\vec{k}-\vec{p}}(\eta_v)+Z_{\vec{k}-\vec{p}}(\eta_u)\overline{Z_{\vec{k}-\vec{p}}(\eta_v)}\bigg)\bigg\}. 
\end{eqnarray}

A similar treatment of the remaining terms in Eq. (\ref{pstate4}) allows the function $\mathfrak{X}_{\vec{k}}^{P_{\eta(t)}}$ in Eq. (\ref{frakxp}) to be written as 
\begin{eqnarray}
\mathfrak{X}_{\vec{k}}^{P_{\eta(t)}} = \mathfrak{X}_{\vec{k}}^0 &+& \frac{\kappa^2}{|X_{\vec{k}}(\eta)|^2}\mbox{ }\int_{\eta_i}^{\eta_f}d\eta_u\int_{\eta_i}^{\eta_f}d\eta_v\mbox{ }\bigg\{h(\eta_u)h(\eta_v)\mbox{ }\mathfrak{X}_{\vec{k}}^{\eta(t)}\bigg(\overline{X_{\vec{k}}(\eta)}X_{\vec{k}}(\eta_u)+X_{\vec{k}}(\eta)\overline{X_{\vec{k}}(\eta_u)}\bigg)\nonumber\\
& &\times\mbox{ }\bigg[\mbox{ }\mathscr{X}_{\vec{k}}^{-+}\bigg(\overline{X_{\vec{k}}(\eta_v)}X_{\vec{k}}(\eta)+X_{\vec{k}}(\eta_v)\overline{X_{\vec{k}}(\eta)}\bigg)G_{\vec{k}}^{-+}(\eta_u,\eta_v)\nonumber\\
& &\mbox{ }-\mbox{ }\mathscr{X}_{\vec{k}}^{++}\bigg(\overline{X_{\vec{k}}(\eta_v)}X_{\vec{k}}(\eta)+X_{\vec{k}}(\eta_v)\overline{X_{\vec{k}}(\eta)}\bigg)G_{\vec{k}}^{++}(\eta_u,\eta_v)\nonumber\\
& &\mbox{ }+\mbox{ }\mathscr{X}_{\vec{k}}^{+-}\bigg(\overline{X_{\vec{k}}(\eta_v)}X_{\vec{k}}(\eta)+X_{\vec{k}}(\eta_v)\overline{X_{\vec{k}}(\eta)}\bigg)G_{\vec{k}}^{+-}(\eta_v,\eta_u)\nonumber\\
& &\mbox{ }-\mbox{ }\mathscr{X}_{\vec{k}}^{--}\bigg(\overline{X_{\vec{k}}(\eta_v)}X_{\vec{k}}(\eta)+X_{\vec{k}}(\eta_v)\overline{X_{\vec{k}}(\eta)}\bigg)G_{\vec{k}}^{--}(\eta_v,\eta_u)\bigg]\bigg\}.
\end{eqnarray}
Here, the exact form of the perturbed state first requires the form of $\mathfrak{X}_{\vec{k}}^0$, $\mathscr{Y}_{\vec{p}}$, and $\mathscr{Z}_{\vec{k}-\vec{p}}$. In order to carry out a numerical calculation these functions may simply be specified to be that of the SLE vacuum or, for example, a Bulk-to-Boundary Approximate KMS state. On the other hand, additional interactions may be considered such that a system of coupled equations for $\mathfrak{X}_{\vec{k}}^0$, $\mathscr{Y}_{\vec{p}}$, and $\mathscr{Z}_{\vec{k}-\vec{p}}$ may be employed. In addition, $\mathscr{E}_{A}^{P_{\eta(t)}}$, as a function of $H_t$, is of course subject to the so called back--reaction problem via the semiclassical Friedmann equation
\begin{equation}
H_t^2 = \frac{8\pi G}{3}\mbox{ }\omega_t^P\bigg(\mbox{T}_{00}(:A(f)^2:)\bigg).
\end{equation}
However, may still in principal carry out a concrete numerical calculation by imposing the solution for $a_t$, i.e. we may take $a_t$ to maintain a fixed form of $a_t^{(\Lambda)}\propto\mbox{ exp}(H_{\Lambda}t)$, $a_t^{(r)}\propto(t-t_i)^{1/2}$, or $a_t^{(m)}\propto(t-t_i)^{3/2}$ during an epoch dominated by a constant vacuum energy($\Lambda$), radiation($r$), or matter($m$) respectively.

\section{Discussion\label{DISC}}

In this work we began with first principles of algebraic quantum field theory in curved spacetime where we employed both the SLE construction of renormalizable ground states and a two-parameter family of automorphisms, including a time averaged perturbation, in describing the dynamics of a dense environment of interacting quantum fields in FRW spacetimes. We then derived for the first time Eq. (\ref{mainres}) as an expression that is in principal amenable to a numerical calculations for the renormalized energy density of a massive, minimally coupled free scalar field perturbed during a finite time interval via quantum interactions, including those far-from-equilibrium, while propagating in a non-stationary spacetime background. This algebraic expression is thus appropriate for computing cosmological observables, i.e. relic abundance calculations associated with common proposals for quantum matter production in the early universe, in order to determine if there are disparities between the algebraic approach and the general approximation, that are in principle experimentally verifiable by future high-precision electromagnetic and/or gravitational-wave detectors. If there are indeed discernible disparities they may serve to illuminate the interplay between quantum interactions and the dynamics of classical spacetime.

An additional application of the algebraic state containing the perturbation derived in Eq. (\ref{frakxp}) is a search for finite time and density corrections to the standard calculation of the observable power spectrum of super-Hubble fluctuations of the proposed quantum field responsible for inflation. Beginning with the linearized Einstein--Klein--Gordon system these fluctuations may be quantized according to the algebraic framework. The gauge invariant perturbations of the field, and hence the comoving curvature perturbations, may then be given the standard treatment via the Bardeen potentials and the Mukhanov-Sasaki variable, \textit{i.e.} a Klein--Gordon field with time-dependent mass \cite{Hac15}. An examination of the spectrum found via the perturbed two-point function may then be compared to that of the spectrum computed in the Bunch--Davies vacuum state. Furthermore, corrections arising from this perturbed algebraic calculation may be probed by a direct comparison with calculations carried out in an effective field theory approach to the operator framework in Ref. \cite{Boy15}. We leave this for future work.

\acknowledgments
We are grateful to R.J. Scherrer for helpful discussions.
\pagebreak


\begin{thebibliography}{99}
\bibitem{KnT} E.W. ~Kolb and M.S. ~Turner, \emph{The Early Universe}, Westview Press (1990)
\bibitem{Wei08} S. ~Weinberg, \emph{Cosmology}, Oxford University Press, Inc., New York (2008)
\bibitem{Bir82} N.D. ~Birrel and P.C. W. ~Davies, \emph{Quantum Fields in Curved Space}, Cambridge University Press, Cambridge U.K. (1982) 
\bibitem{Par09} L.E. ~Parker and D.J. ~Toms, \emph{Quantum Field Theory in Curved Spacetime}, Cambridge University Press, Cambridge U.K. (2009) 
\bibitem{LeB96} M. ~Le Bellac \emph{Thermal Field Theory}, Cambridge University Press, Cambridge U.K. (1996) 
\bibitem{Das97} A. ~Das, \emph{Finite Temperature Field Theory}, World Scientific Publishing Co., Pte. Ltd. (1997) 
\bibitem{Boy05} D. ~Boyanovsky, K. ~Davey and C.M. ~Ho, Phys. Rev. D \textbf{71} 023523  (2005) \href{http://arxiv.org/abs/hep-ph/0411042}{[hep-ph/0411042v2]}.
\bibitem{Ani08} A. ~Anisimov, W. ~Buchmuller, M. ~Drews, and S. ~Mendizabal, Annals Phys. \textbf{324} 1234 (2009) \href{http://arxiv.org/abs/0812.1934v2}{[hep-ph/0812.1934v2]}.
\bibitem{Ham12} K. ~Hamaguphi, T. ~Moroi,  and K. ~Mukaida, JHEP \textbf{1201} 083 (2012)  \href{http://arxiv.org/abs/1111.4594v2}{[hep-ph/1111.4594v2]}.
\bibitem{HoS15} C.M. ~Ho and R.J. ~Scherrer, Phys. Rev. D \textbf{92} 025019 (2015) \href{http://arxiv.org/abs/1503.03534v2}{[hep-ph/1503.03534v2]}.
\bibitem{Dre16} M. ~Drewes and J.U. ~Kang, JHEP \textbf{1605} 051 (2016) \href{http://arxiv.org/abs/1510.05646v3}{[hep-ph/1510.05646v3]}.
\bibitem{Bru03} R. ~Brunetti, K. ~Fredenhagen, and R. ~Verch, Commun. Math. Phys. \textbf{237}  31 (2003) \href{https://arxiv.org/abs/math-ph/0112041}{[math-ph/0112041]}.
\bibitem{Wal94} R.M. ~Wald, \emph{Quantum Field Theory in Curved Spacetimes and Black Hole Thermodynamics}, The University of Chicago Press (1994) 
\bibitem{Kha15} I. ~Khavkine and V. Moretti, \emph{Algebraic QFT in curved spacetime and quasifree Hadamard states: an introduction} in Ch. 5 of \emph{Advances in Algebraic Quantum Field Theory}, edited by R. Brunetti et al., Springer International Publishing (2015)  \href{http://arxiv.org/abs/1412.5945v3}{[math-ph/1412.5945v3]}
\bibitem{Deg13} A. ~Degner, \emph{Properties of states of low energy
on cosmological spacetimes}, Ph.D. Thesis, University of Hamburg (2013) \href{https://www.physnet.uni-hamburg.de/services/biblio/dissertation/dissfbPhysik/___Volltexte/Andreas___Degner/Andreas___Degner.pdf}{[A. Degner Ph.D. Thesis]}.
\bibitem{Hac15} T.P. ~Hack, \emph{Cosmological applications of algebraic quantum field theory in curved spacetimes}, Springer Briefs in Mathematical Physics \textbf{6} (2016) \href{https://arxiv.org/abs/1506.01869v1}{[gr-qc/1506.01869v1]}.
\bibitem{Hol01} S. ~Hollands and R.M. ~Wald, Commun. Math. Phys. \textbf{223} 289 (2001) \href{https://arxiv.org/abs/gr-qc/0103074}{[gr-qc/0103074v2]}
\bibitem{Hol02} S. ~Hollands and R.M. ~Wald, Commun. Math. Phys. \textbf{231} 309 (2002) \href{https://arxiv.org/abs/gr-qc/0111108v2}{[gr-qc/0111108v2]}.
\bibitem{Mor03} V. ~Moretti, Commun. Math. Phys. \textbf{232} 189 (2003) \href{https://arxiv.org/abs/gr-qc/0109048}{[gr-qc/0109048v2]}.
\bibitem{Hol05} S. ~Hollands and R.M. ~Wald, Rev. Math. Phys. \textbf{17} 227 (2005) \href{https://arxiv.org/abs/gr-qc/0404074}{[gr-qc/0404074v2]}.
\bibitem{Oji86} I. ~Ojima, \emph{Quantum field theoretical approach to non-equilibrium dynamics in curved spacetime} (1986) \href{http://inspirehep.net/record/227022}{[inspirehep.net:227022]} 
\bibitem{Buc02} D. ~Buchholz, J. ~Mund, and S.J. ~Summers, Clas. Quant. Grav. \textbf{19} 6417 (2002) \href{https://arxiv.org/abs/hep-th/0207057v1}{[hep-th/0207057v1]}.
\bibitem{Par69} L. ~Parker, Phys. Rev. \textbf{183} 1057 (1969) \href{https://journals.aps.org/pr/abstract/10.1103/PhysRev.183.1057}{[PhysRev:183.1057]}
\bibitem{Lue90} C. ~Lueders and J.E. ~Roberts, Commun. Math. Phys. \textbf{134} 29 (1990) \href{https://projecteuclid.org/download/pdf_1/euclid.cmp/1104201612}{[euclid.cmp:1104201612]}
\bibitem{Haa84} R. ~Haag, H. ~Narnhofer, and U. ~Stein, Commun. Math Phys. \textbf{94} 219 (1984) \href{https://link.springer.com/article/10.1007/BF01209302}{[springer.com/article/10.1007/BF01209302]}.
\bibitem{Few00} C.J. ~Fewster, Class. Quant. Grav. \textbf{17} 1897 (2000) \href{https://arxiv.org/abs/gr-qc/9910060}{[gr-qc/9910060v2]}.
\bibitem{Olb07} H. ~Olbermann, Class. Quant. Grav. \textbf{24} 5011 (2007) \href{https://arxiv.org/abs/0704.2986}{[gr-qc/0704.2986v2]}.
\bibitem{Deg10} A. ~Degner and R. Verch, J. Math. Phys. \textbf{51} 022302 (2010) \href{https://arxiv.org/abs/0904.1273v1}{[gr-qc/0904.1273v1]}.
\bibitem{Chi08} B. ~Chilian and K. Fredenhagen, Commun. Math. Phys. \textbf{287} 513 (2009) \href{https://arxiv.org/abs/0802.1642v3}{[math-ph/0802.1642v3]}.
\bibitem{Buc02S}  D. ~Buchholz, I. ~Ojima, and H. ~Roos, Annals Phys. \textbf{297} 219 (2002) \href{https://arxiv.org/abs/hep-ph/0105051}{[hep-ph/0105051v2]}.
\bibitem{Oji03} I. ~Ojima, \emph{How to formulate non-equilibrium local states in QFT: general characterization and extension to curved spacetime} in \emph{A Garden of Quanta} edited by J. Arafune et al., World Scientific Publishing Co., Pte. Ltd. (2003) \href{https://arxiv.org/abs/cond-mat/0302283v1}{[cond-mat.stat-mech/0302283v1]}.
\bibitem{Gra16} M. ~Gransee, \emph{Local thermal equilibrium states in relativistic quantum field theory} in \emph{Quantum Mathematical Physics: A Bridge between Mathematics and Physics} edited by F. Finster et al.,  Springer International Publishing (2016) \href{https://arxiv.org/abs/1602.09110v1}{[math-ph/1602.09110v1]}.
\bibitem{Sch61} J. S. ~Schwinger, J. Math. Phys. \textbf{2} 407 (1961) \href{http://aip.scitation.org/doi/abs/10.1063/1.1703727}{[aip.scitation.org:1.1703727]} 
\bibitem{Kel64} L. ~Keldysh, Sov. Phys. -- JETP \textbf{20} 4 (1964) \href{http://www.jetp.ac.ru/cgi-bin/e/index/e/20/4/p1018?a=list}{[jetp.ac.ru:1018]}  
\bibitem{Pla15p} P.A.R. ~Ade and others, \emph{Planck 2015 results. XIII. Cosmological parameters} (2015) \href{http://arxiv.org/pdf/1502.01589v2.pdf}{[astro-ph/1502.01589]]}.
\bibitem{Pla15i} P.A.R. ~Ade and others, \emph{Planck 2015 results. XX. Constraints on inflation} (2015) \href{http://arxiv.org/pdf/1502.02114v1.pdf}{[astro-ph/1502.02114]}.
\bibitem{Hol06} S. ~Hollands, Commun. Math. Phys. \textbf{273} 1 (2006) \href{https://arxiv.org/abs/gr-qc/0605072v1}{[gr-qc/0605072v1]}.
\bibitem{Hol08a} S. ~Hollands, Rev. Math. Phys. \textbf{20} 1033 (2008) \href{https://arxiv.org/abs/0705.3340v4}{[gr-qc/0705.3340v4]}.
\bibitem{Hol08b} S. ~Hollands and R.M. ~Wald, Gen. Rel. Grav. \textbf{40} 2051 (2008) \href{https://arxiv.org/abs/0805.3419v1}{[gr-qc/0805.3419v1]}.
\bibitem{Hol10} S. ~Hollands and R.M. ~Wald, Commun. Math Phys. \textbf{293} 85 (2010) \href{https://arxiv.org/abs/0803.2003v1}{[gr-qc/0803.2003v1]}.
\bibitem{Sch08} J. ~Schlemmer and R. ~Verch, Ann. Heni Poincar\'{e} \textbf{9} 1945 (2008) \href{https://arxiv.org/abs/0802.2151v1}{[gr-qc/0802.2151v1]}.
\bibitem{Dap09a} C. ~Dappiaggi, V. ~Moretti, and N. ~Pinamonti, Commun. Math. Phys. \textbf{285} 1129 (2009) \href{https://arxiv.org/abs/0712.1770v3}{[gr-qc/0712.1770v3]}.
\bibitem{Dap09b} C. ~Dappiaggi, V. ~Moretti, and N. ~Pinamonti, J. Math. Phys. \textbf{50} 062304 (2009) \href{https://arxiv.org/abs/0812.4033}{[gr-qc/0812.4033]}.
\bibitem{San10} K. ~Sanders, Commun. Math. Phys. \textbf{295} 498 (2010) \href{https://arxiv.org/abs/0903.1021v1}{[math-ph/0903.1021v1]}.
\bibitem{Lin13} F. ~Lindner, \emph{Perturbative algebraic quantum field theory at finite temperature}, Ph.D. Thesis, University of Hamburg (2013) \href{http://www-library.desy.de/preparch/desy/thesis/desy-thesis-13-029.pdf}{[F. Lindner Ph.D. Thesis]}
\bibitem{Fre14} K. ~Fredenhagen and F. ~Lindner, Commun. Math. Phys. \textbf{332} 895 (2014) \href{https://arxiv.org/pdf/1306.6519.pdf}{[math-ph/1306.6519v5]}
\bibitem{Hac18} T.P. ~Hack and R. Verch, \emph{Non-equilibrium steady states for the interacting Klein--Gordon
field in 1+3 dimensions} (2018) \href{https://arxiv.org/pdf/1806.00504.pdf}{[math-ph/1806.00504v1]} 
\bibitem{Kus08} M. ~K{\"u}sk{\"u}, \emph{A Class of Almost Equilibrium States in Robertson-Walker Spacetimes}, Ph.D. Thesis, University of Hamburg (2008) \href{https://arxiv.org/abs/0901.1440}{[hep-th/0901.1440]}.
\bibitem{Dap11} C. ~Dappiaggi, T.P. ~Hack, and N. ~Pinamonti, Ann. Heni Poincar\'{e} \textbf{12} 1449 (2011) \href{https://arxiv.org/abs/1009.5179}{[gr-qc/1009.5179]}.
\bibitem{Boy15} D. ~Boyanovsky, New J. Phys. \textbf{17} 063017 (2015) \href{http://arxiv.org/abs/1503.00156v2}{[hep-ph/1503.00156v2]}.
\end{thebibliography}
\end{document}